\documentclass[journal,9pt]{IEEEtran}

\usepackage{amsthm}
\usepackage{ragged2e}
\usepackage[utf8]{inputenc}
\usepackage{amsmath,amssymb,amsmath,amsthm,bm,graphicx,xcolor}
\usepackage[breaklinks]{hyperref}
\usepackage{color}
\usepackage{amssymb}
\usepackage{graphicx}
\usepackage{xcolor}
\usepackage{float}
\usepackage{placeins}
\usepackage{lipsum}
\usepackage{subfigure}
\hypersetup{colorlinks=true}
\usepackage{mathtools}

\usepackage{caption} 
\usepackage{multirow}            
\usepackage{makecell}
\definecolor{redX}{HTML}{D40200} 
\definecolor{violetX}{HTML}{9D1569}
\begin{document}

\title{Topology optimization of a superabsorbing thin-film semiconductor metasurface}
\author{Johannes Gedeon, Izzatjon Allayarov, Emadeldeen Hassan, and Antonio Cal{\`a} Lesina
\thanks{
This work was supported by the Deutsche Forschungsgemeinschaft (DFG, German Research Foundation) under Germany’s Excellence Strategy within the Cluster of Excellence PhoenixD (EXC 2122, Project ID 390833453), and within the Research Grant CA 2763/2-1 (Project ID 527470210). The computing time was granted by the Resource Allocation Board and provided on the supercomputer Lise and Emmy/Grete at NHR@ZIB and NHR@Göttingen as part of the NHR infrastructure. The calculations for this research were conducted with computing resources under the project nip00059.}
\thanks{Johannes Gedeon, Izzatjon Allayarov and Antonio Cal{\`a} Lesina are with
the Hannover Centre for Optical Technologies, the Institute for Transport and Automation Technology (Faculty of Mechanical Engineering), and the Cluster of Excellence PhoenixD,
Leibniz University Hannover, 30167 Hannover, Germany (e-mail: johannes.gedeon@hot.uni-hannover.de; antonio.calalesina@hot.uni-hannover.de).}
\thanks{Emadeldeen Hassan is with the Department of Applied Physics and Electronics, Umeå University, SE-901 87 Umeå, Sweden (e-mail: emadeldeen.hassan@umu.se).}}


\maketitle

\begin{abstract}
We demonstrate a computational inverse design method for optimizing broadband-absorbing metasurfaces made of arbitrary dispersive media. 
Our figure of merit is the time-averaged instantaneous power dissipation in a single unit cell within a periodic array. Its time-domain formulation allows capturing the response of arbitrary dispersive media over any desired spectral range. Employing the time-domain adjoint method within a topology optimization framework enables the design of complex metasurface structures exhibiting unprecedented broadband absorption.
We applied the method to a thin-film Silicon-on-insulator configuration and explored the impact of structural and (time-domain inherent) excitation parameters on performance over the visible–ultraviolet. \textcolor{black}{We provide a physical insight into the dissipation mechanism of the optimized structures.} Since our incorporated material model can represent any linear material, the method can also be applied to other all-dielectric, plasmonic, or hybrid configurations.
\end{abstract}

\begin{IEEEkeywords}
Absorption, adjoint method, complex-conjugate pole–residue pairs model, FDTD method, inverse design, metasurface, optical dispersion, quasi-guided modes, silicon, surface lattice resonances, time domain, topology optimization
\end{IEEEkeywords}

\section{Introduction}\label{Sec:Introduction}

Thin-film absorbing metasurfaces are of great interest for both fundamental research and engineering, as they can efficiently capture light across a broad spectrum while reducing material resource use. Among the various potential applications, they attract the most attention in the field of solar energy harvesting~\cite{Mascaretti2023, EnergyConversion, Chang2018}. To date, numerous studies are being conducted to improve light trapping in \mbox{(nano-)} photonic devices to achieve broadband absorption, and different architectures and material combinations have been proposed. {\color{black}These include metacoatings exhibiting graded refractive index proﬁles~\cite{Chen2024}, hybrid metal-dielectric–metal stacks and tapered structures for surface plasmon coupling~\cite{Zhu2012, stack, Li2014, KarimiHabil2022}, and supercells that achieve improved multi- and broadband control \cite{Nielsen:12, Nagarajan2018, SolarAbsorber}, to name just a few.}

\textit{All-dielectric} absorbers, however, rely on complexly shaped resonators that exhibit multiple resonances and their efficient coupling~\cite{Pala2016, Mitrofanov2020}. Pala et al.~\cite{Pala2016} demonstrated that using waveguide geometries as subwavelength elements can enhance light trapping over a broad bandwidth. As experimentally proved, their thin-film designs exhibit a remarkable increase in short-circuit current, making them attractive for photovoltaic applications. This boost in absorption comes from a combination of localized Mie resonances and delocalized resonant waveguide modes. While Mie resonances emerge in single meta-atoms \cite{GedeonDissip}, delocalized modes originate from collective effects due to the array configuration of a metasurface. These include surface lattice resonances~(SLRs) and the above-mentioned (quasi)~guided modes~(QGMs)~\cite{Castellanos2019}. SLRs arise from the coupling of resonances in meta-atoms (which need not be physically interconnected) by the in-plane diffraction orders of the lattice. They appear when the diffraction order becomes an evanescent wave, known as the Rayleigh anomaly~(RA). QGMs are commonly not distinguished from SLRs in the literature; however, they require a waveguide structure with a higher refractive index than the surrounding medium. In Ref.~\cite{Pala2016}, such QGMs were employed in periodically arranged ridge and trapezoidal waveguide structures to boost broadband absorption; {\color{black}these structures were found based on prior physical insights and parametric sweeps of geometrical features.}

{\color{black}Conventional design approaches rely on dimensional variations of simple geometries~\cite{Li2014, Mitrofanov2020}, additional prior analysis of such single subelements for building supercells~\cite{Nagarajan2018}, predetermined (stack) combinations of different media~\cite{Chen2024, stack, KarimiHabil2022}, or physical insight/intuition to build more complex topologies~\cite{Pala2016}. These approaches can be time-consuming and computationally expensive.} In recent years, \textit{inverse design} methods have emerged as a powerful alternative, where the structural design is dictated by the desired optical response itself~\cite{Li2022}. However, broadband optimization still poses significant challenges for methods developed in the frequency domain~\cite{Hammond:22}. In a recent work~\cite{GedeonDissip}, we introduced a novel gradient-based inverse design method employing topology optimization~(TopOpt) to enhance broadband absorption in dispersive nanoparticles. Our approach leverages the adjoint method in the \textit{time domain} to efficiently compute gradients over any desired spectral range, enabling the design to evolve into a structure that inherently accounts for frequency-dependent material responses over a broad spectrum. This method promises to be efficient for designing thin-film all-dielectric absorbers as well. {\color{black}
In contrast to previous design approaches, our structural optimization method for maximizing the (local) power dissipation allows for an unbiased exploration over the entire topological (3D)~design space via a gradient-based updating routine; it further provides for a flexible change of the design media by simply switching the material parameters as the input, and does not require explicit evaluation of the absorption spectra over broadband range during the optimization process.} 

In this article, we demonstrate our TopOpt method for the design of a superabsorbing Silicon metasurface in the visible–ultraviolet range. Our method allows us to optimize a metasurface beyond the limitations of conventional geometries, resulting in unprecedented topologies and phenomena that can also inspire other (conventional) design approaches. 
To explore the theoretical limits of broadband absorption, we enable free-form optimization in 3D; and in this context, we provide physical insight into both improvements and limitations in achieving uniform absorption.

\section{Complex–conjugate pole-residue pairs and instantaneous power dissipation in dispersive media}\label{Sec:Method}
\begin{figure}[ht!]
        \centering
        \includegraphics[trim = 0mm 0mm 0mm 0mm,width=0.9\linewidth]{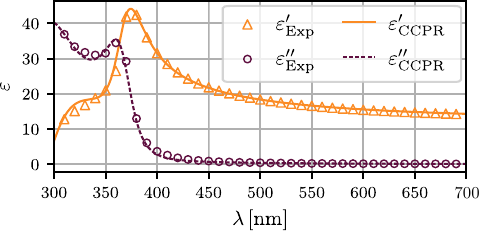}
        \caption{%
      \label{Fig:EpsSilicon}
      \justifying Real and imaginary part of the permittivity \mbox{$\varepsilon = \varepsilon^{\prime} - j\,\varepsilon^{\prime\prime}$} of Silicon, reproduced by the CCPR model~(\ref{Eq:eps_CCPR}) using the coefficients listed in Table~\ref{tab:epsilonSi}. The experimental data was taken from Ref.~\cite{Si_Schinke}.
    }
\end{figure}

To enable the inverse design of absorbing metasurfaces, we build on our recent contributions from Refs.~\cite{Gedeon2023, GedeonDissip}. Herein, we introduced a time-domain adjoint scheme for the design of nanostructures of arbitrary, dispersive media, which are described by the complex–conjugate pole residue~(CCPR) pairs model~\cite{CCPRmodel},
\begin{equation}\label{Eq:eps_CCPR}
\varepsilon(\omega)=\varepsilon_{\infty}+\frac{\sigma}{j \omega \varepsilon_0}+\sum_{p=1}^{P}\left(\frac{c_{p}}{j \omega-a_{p}}+\frac{c_{p}^*}{j \omega-a_{p}^*}\right).
\end{equation}
The coefficients $a_p$ and $c_p$ can be complex (with~$*$ denoting complex conjugation). Selecting the number of poles $P$ appropriately allows the model to accurately fit the complex permittivity of a linear material over an arbitrary frequency range. 
The coefficients for fitting Silicon’s permittivity in the range \mbox{300–700 nm} are given in Table~\ref{tab:epsilonSi}, with only two poles used. Fig.~\ref{Fig:EpsSilicon} demonstrates the excellent agreement between the modeled complex permittivity and the experimental measurements. 
\begin{table}[ht!]
    \centering
    \begin{tabular}{l c}
        \hline
        Parameter & Fitting value \\ 
        \hline 
        $\varepsilon_{\infty}$ & $1$ \\[2pt]
        $\sigma$               & $0$ \\
        $a_1$                  & $-8.00 \times 10^{14}+6.39 \times 10^{15} j$ \\
        $c_1$                  & $\phantom{-}7.31 \times 10^{14}-2.89 \times 10^{16} j$ \\
        $a_2$                  & $-2.32 \times 10^{14}+5.12 \times 10^{15} j$ \\
        $c_2$                  & $\phantom{-}4.68 \times 10^{15}-4.55 \times 10^{15} j$ \\[2.5pt]
        \hline
    \end{tabular}
    \caption{CCPR parameters for modeling Silicon's permittivity $\varepsilon$ in the wavelength range~\mbox{300-700\,nm}, using two poles.}
    \label{tab:epsilonSi}
\vspace{-10pt}    

\end{table}

We integrated this model into the time-domain Maxwell's equations by employing the auxiliary differential equations~(ADEs) method \cite{HanCCPRADEs}. It allows us to accurately capture the material response over the spectral range covered in the time signal of excitation. These equations formed the basis for our derived adjoint scheme to enable gradient-based TopOpt~\cite{TopOptGeneral} that is used to find an optimal material distribution that maximizes a specified objective. Due to the universality of the CCPR model, the framework provides maximum flexibility when handling dispersive media for broadband optimization. Furthermore, we outlined how the algorithm can be incorporated into the Finite-Difference Time-Domain~(FDTD) method - a widely used computational tool for modeling electrodynamic phenomena \cite{Taflove}.

Dispersion is inherently linked to material losses \cite{landau1995electrodynamics}, which cause power dissipation in dispersive nanostructures. In our recent work~\cite{GedeonDissip}, we found that the \textit{instantaneous power dissipation density} based on the CCPR-ADE formulation reads
\begin{equation}\label{Eq:InstantanousDissipation}
q_e(t) = \sigma \mathbf{E}^{2} + 2\sum_{p=1}^P \Re\left\{\frac{(\partial_{t}\mathbf{Q}_p)^2}{\varepsilon_{0}c_p}\right\},
\end{equation}
where $\mathbf{E}$ is the electric field, and $\mathbf{Q}_p$ denotes the auxiliary field associated with a CCPR pole. 
The time-averaged loss in the nanostructure can then be calculated using
\begin{equation}\label{Eq:ObjectiveDissipation}
F := \frac{1}{T}\int_{\Omega_{\mathrm{o}} \times I} q_{e}(t,\bm{r}) \,\mathrm{d} t\mathrm{d}^3r,   
\end{equation}
where $I=[0, T]$ is a time interval, and $\Omega_{\mathrm{o}}$ represents the observation region. Given the spectral content conveyed by the excitation time signal, it accounts for the absorptive response across the entire frequency range. This expression serves as the objective function for the metasurface optimization presented in this article. 
{\color{black} A brief introduction to the CCPR-ADE formalism, the derivation of the instantanous power dissipation density~(\ref{Eq:InstantanousDissipation}), and a summary of our derived (FDTD) adjoint method for optimization are given in Sections~I-III in the Supplementary Material.}

\section{Setup}\label{Sec:Setup}

\begin{figure}[ht]
\centering
\includegraphics[trim = 0mm 0mm 0mm 3mm,width=1\linewidth]{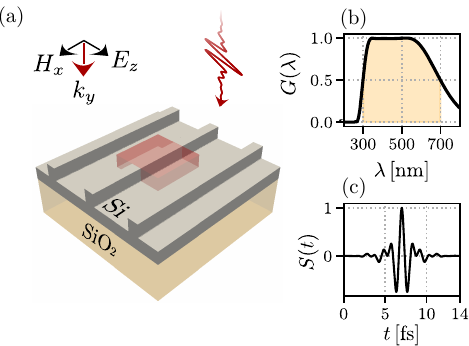}
\caption{\label{Fig:Setup} \justifying Sketch of the optimization setup. (a)~The red cuboid marks our design region and represents a single unit cell of the periodic Silicon~(Si) metasurface, enclosed by SiO$_{2}$ and air. The system is excited by a light pulse, covering the (b)~\mbox{wavelengths 300-700 nm} (marked in yellow) through the (c)~time signal.}
\end{figure}

For the demonstration of our method, we choose a thin-film Silicon-on-insulator configuration as in Fig.\,\ref{Fig:Setup}(a). The design domain coincides with a single unit cell of an infinitely extended periodic metasurface, surrounded by air and SiO$_{2}$. Here, the Silicon metasurface is shown as a ridge structure only for illustration purposes. This conventional design studied in Ref.~\cite{Pala2016} is used as a benchmark to evaluate the performance of our actual TopOpt design. We consider a constant refractive index for air~($n=1$) and SiO$_{\text{2}}$~(n=1.45), and the permittivity of Silicon $\varepsilon_{\text{Si}}$ modeled by the CCPR~(\ref{Eq:eps_CCPR}) with its coefficients given in Table~\ref{tab:epsilonSi}.

We aim to maximize the absorption of power delivered by a $z$-polarized plane wave at normal incidence over the wavelength range~\mbox{300–700\,nm}. This is achieved by maximizing the time-averaged instantaneous power dissipation, Eq.~(\ref{Eq:ObjectiveDissipation}), integrated over the unit cell, while imposing periodic boundary conditions~(PBC). The time~$T$ in Eq.~(\ref{Eq:ObjectiveDissipation}) is chosen such that the time-domain fields are sufficiently decayed and no longer contribute to a substantial increase in the objective. During the optimization, the system is excited by a short pulse governed by a tailored time signal. This time signal covers the entire spectral range of interest, as shown in Fig.\,\ref{Fig:Setup}(b)\,and\,(c).

We employ a density-based TopOpt~\cite{TopOptGeneral} and our adjoint formulation \cite{Gedeon2023} to find an optimal material distribution that maximizes the absorption over broad range. The optimization algorithm iteratively updates the design in the form of a density distribution $\rho(\bm{r})\in[0,1]$ within the unit cell. The density represents a linear interpolation of both permittivities of air and Silicon, \mbox{$\varepsilon_{\text{Mix}}=\rho\,\varepsilon_{\text{Si}} + (1-\rho)\, \varepsilon_{\text{Air}}$ }. We note that $\varepsilon_{\text{Mix}}$ is frequency-dependent, but recall that due to our time-domain adjoint formulation using CCPR-ADEs, the algorithm inherently accounts for dispersion and material loss across the entire spectral range of interest. 
As the optimization progresses, the density distribution gradually converges towards a binary structure, where values of $\rho=1$ correspond to the design material and $\rho=0$ to the background material. At the end of the optimization, the density is binarized based on a threshold to obtain the final structure.  We then evaluate the broadband performance of the designed metasurface by analyzing its absorption spectrum (using \textit{Lumerical's}~\texttt{FDTD}~solver~\cite{LumericalFDTD}).

All following optimizations are conducted using an in-house parallelized FDTD~solver~\cite{Lesina:15} with TopOpt implementation. The FDTD and TopOpt configuration settings are adapted from the Silicon particle optimization described in Ref.~\cite{Gedeon2023}. All presented designs are openly available (as STL format)~\cite{dataset} and can be integrated into an electromagnetic solver to validate our results. In the following, we will explicitly specify optimization parameters of only significant importance. Moreover, if we refer to ``the design'' (or structure) in our descriptions, it is always meant in the context of its periodic arrangement throughout this article.

\section{Results and Discussion}\label{Sec:Opt}
\subsection{A first demonstration}
We first apply our method to the design of a Silicon metasurface of thickness~\mbox{$d=200$\,nm}, having a quadratic unit cell with periodicity~\mbox{$a=400$\,nm}. With the discretization we chose within the FDTD method, this results in $1.5\times 10^{6}$ degrees of freedom of the design to evolve. The performance is compared with a conventional Ridges structure as reported in Ref.~\cite{Pala2016} as a reference. Our TopOpt configuration allows the evolution to metasurfaces exploiting SLRs or QGMs as collective effects to enhance absorption, which were briefly explained in the introduction. To maintain generality, we hereafter refer to SLRs and QGMs collectively as lattice resonances~(LRs), given that both phenomena can emerge in principle and both are dictated by the Rayleigh anomaly~(RA).

\begin{figure}[ht!]
        \centering
        \includegraphics[trim = 0mm 0mm 0mm 0mm, clip ,width=1\linewidth]{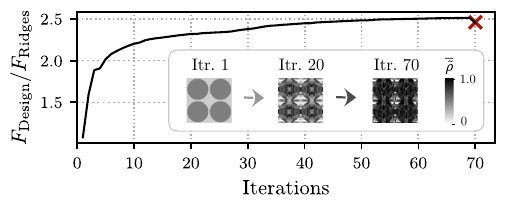}
        \caption{\justifying Convergence plot of the optimization. $F_{\text{Design}}$ is the time-averaged measured power dissipation of the evolving structure, normalized to that of a conventional Ridges~($F_{\text{Ridges}}$). The subfigure presents top views of the (filtered and projected~\cite{Wang2011OnPM}) density $\bar{\tilde{\rho}}$ arranged in a~$2\times2$ array, capturing the metasurface's evolution. At the last iteration, a thresholding was applied to yield the binary design (marked by a red cross).}
        \label{Fig:Convergence}  
\end{figure}

As the initial density, we choose a cylinder~\mbox{($\rho=0.5$)} embedded in a homogeneous environment ~\mbox{($\rho=0.4$)} that fills the entire unit cell. Fig.\,\ref{Fig:Convergence} shows the convergence of the optimization over 70~iterations, tracking the objective in Eq.~(\ref{Eq:ObjectiveDissipation}) for the evolving structure normalized to that of the Ridges. The density steadily progressed toward a binary design. After applying the final thresholding, we obtained an enhancement factor~of~$2.45$. We note that starting with a \textit{homogeneous} density across the entire unit cell prevents the design from evolving into a complex 3D~structure, as all fields become in-plane invariant due to our physical setup (Fig.\,\ref{Fig:Setup}). In that case, the converging density exhibits a layered profile similar to that of metacoatings~\cite{Chen2024}.

\begin{figure}[ht!]
\centering
        \includegraphics[trim = 0mm 0mm 0mm 4mm,width=1\linewidth]{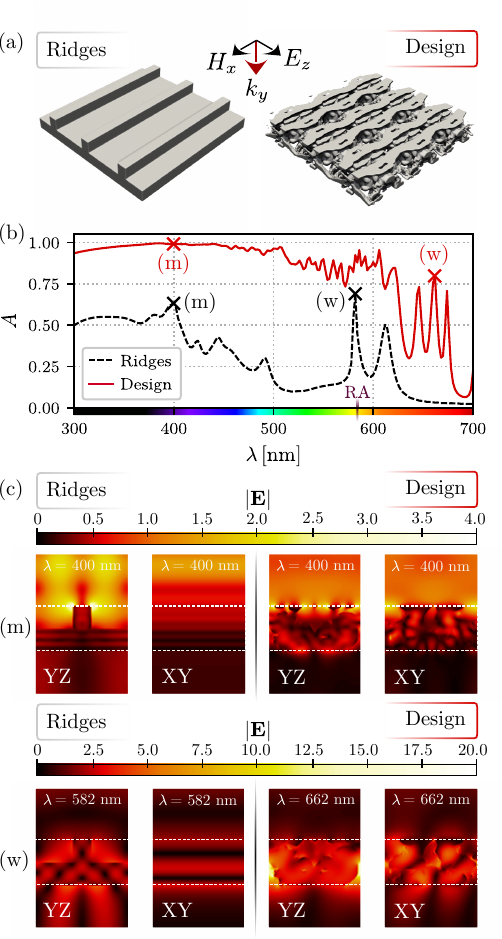}
        \caption{\justifying Comparison of a conventional Ridges design with our TopOpt metasurface from Fig.\,\ref{Fig:Convergence}. (a)~Illustration of both metasurfaces by showing a section of 9 adjacent unit cells, and (b)~their absorption spectrum over the wavelength range of interest. Here, `RA' denotes SiO$_{2}$'s first-order Rayleigh anomaly at \mbox{$\lambda^{\mathrm{RA}}=584$\,nm}. The crosses mark absorption maxima associated with Mie-type resonances~`(m)' and (quasi)~guided modes~`(w)'; and (c)~displays cross-sections of the corresponding electric field's magnitude in a unit cell of both metasurfaces {\color{black}(normalized to the spectrum of the incident pulse)}. The boundaries of the Silicon thin-film are marked as white dashed lines.}
        \label{Fig:FirstResults}
\end{figure}

\begin{figure}[ht!]
	\centering
	\includegraphics[width=1\linewidth]{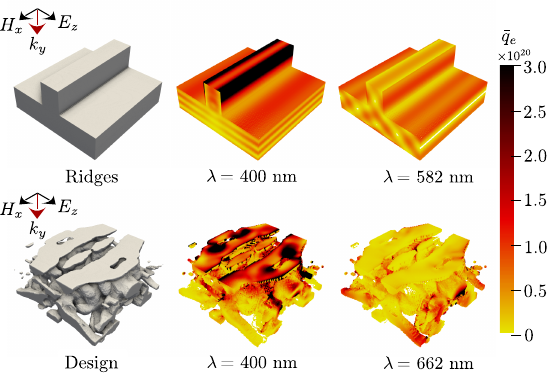}
	\caption{{\color{black}Time-averaged electric power dissipation density, \mbox{$\bar{q}_{e}(\omega)= \frac{1}{2}\varepsilon_{0}\,\omega \,\Im\left\{ \varepsilon(\omega) \right\}|\mathbf{E}|^{2}$}~\cite[\S 61]{landau1995electrodynamics} (with \mbox{$\omega= 2\pi c / \lambda$}), in a single unit cell of both metasurfaces from Fig.\,\ref{Fig:FirstResults}. The wavelengths correspond to those marked by `(m)' and `(w)' in Fig.\,\ref{Fig:FirstResults}(b). The dissipation profile represents the spatial distribution of the fraction of absorbed power, normalized to the power of the incident source (in units $m^{-3}$). The full (3D) data of the time-averaged power dissipation, together with the time-harmonic electric field and Poynting vector, are available in the dataset~\cite{dataset}.}}
	\label{fig:Dissip3D}
\end{figure}

 Interestingly, the periodic arrangement of the unit cell's optimized material distribution forms a waveguide pattern as of the Ridges metasurface{\color{black}, as shown in Fig.\,\ref{Fig:FirstResults}(a)}. We compared their broadband performance by analyzing their absorption spectrum in Fig.\,\ref{Fig:FirstResults}(b). Our TopOpt design shows a remarkable broadband performance, reaching its absorption peak~{\color{black}\mbox{$A_{\text{Max}}=0.995$} at~\mbox{$\lambda=384$\,nm}}. It outperforms the conventional Ridges structure over the entire wavelength range~\mbox{300\,-\,700 nm}, demonstrating the success and strength of the inverse design method. 
 
 {\color{black}
  As it was already observed in Ref.~\cite{Pala2016}, the Ridges exhibit two characteristic absorption peaks around~\mbox{$\lambda=600$\,nm}, which were attributed to coupling with QGMs. Similar features also appear in our TopOpt design in the red region~\mbox{$\lambda>630$\,nm}. We labeled features we associate with localized Mie-type resonances as `(m)' and those we attribute to QGMs as `(w)'. The corresponding cross-sections of the time-harmonic electric field magnitude~\mbox{$|\mathbf{E}|$} in one unit cell are visualized in Fig.\,\ref{Fig:FirstResults}(c). The absorption is determined by the (volume integral) of the time-averaged power dissipation density within the nanostructures, which is illustrated in Fig.\,\ref{fig:Dissip3D}. Comparing the profile associated with localized Mie-type resonances at \mbox{$\lambda=400$\,nm} in both structures~(Fig.\,\ref{Fig:FirstResults}(c), 1st row), we observe enhanced light trapping in the complex, branched TopOpt design compared to the Ridges, characterized by numerous diffusely distributed hotspots (see also Fig.\,\ref{fig:Dissip3D}, comparing 1st and 2nd row for \mbox{$\lambda=400$\,nm}). The enhanced dissipation for the Ridges comes primarily due to the strong localized field enhancement along its (top) outer edges - in contrast to its delocalized mode at \mbox{$\lambda=582$\,nm}~(Fig.\,\ref{Fig:FirstResults}, YZ cross-sections; and Fig.\,\ref{fig:Dissip3D}, 1st row). The electric field distribution corresponding to their characteristic absorption peaks associated with QGMs~(Fig.\,\ref{Fig:FirstResults}(c), 2nd row) resembles a profile typically observed in waveguiding structures~\cite{Huang2023}. A visualization of the power flow in the case of the Ridges for both wavelengths can be found in Section~IV of the Supplementary Material, in which we further motivate the distinction between Mie-type and QGMs resonances.
 }
 
The absorption spectrum of the TopOpt metasurface reveals a performance profile similar to that previously observed in the single-particle optimization reported in Ref.~\cite{GedeonDissip}: the design shows strong performance in the (ultra)~violet regime but declines towards the (infra)~red. The optimizer leverages Silicon’s intrinsic material's losses toward the (ultra) violet. This is the region where interband transitions can occur, and are indicated by an increased imaginary part of the permittivity~$\varepsilon^{\prime\prime}$ in Fig.\,\ref{Fig:EpsSilicon}. In a thin-film–on–insulator configuration (as ours), the semiconductor can even support plasmonic resonances in the ultraviolet \cite{Dong2019} - a phenomenon typically associated with metals. Given the algorithm's gradient-based nature and the specified spectral weighting (Fig.\,\ref{Fig:Setup}(b)), it efficiently directs the design to evolve toward a structure that attains a (local) maximum of the objective Eq.~(\ref{Eq:ObjectiveDissipation}), exhibiting broadband absorption across the inherently prioritized wavelengths. This is the signature trait of the time-domain approach and what fundamentally distinguishes it from the commonly used (hybrid) frequency-domain methods~\cite{Hammond:22}.

{\color{black}As shown in Fig.\,\ref{Fig:FirstResults}(b), our optimization result suggests that the optimizer exploits LRs over a broad range to boost absorption toward the red through the appearance of sharper features in the absorption spectrum around the substrate’s first-order RA at~\mbox{$\lambda^{\mathrm{RA}} = a \cdot n_{\mathrm{SiO}_2} = 584\ \mathrm{nm}$}~\cite{RayleighCond}; due to the evolved topological interconnectivity (see Fig.\,\ref{Fig:FirstResults}(a)), we attribute these effects to the emergence of QGMs.
}
However, LRs' mode density and strength highly depend on periodicity and thin-film thickness \cite{Garcia2021}, and likely on the induced spectral weighting of the time signal under which the design evolves.  Therefore, we examine the influence of structural and excitation parameters on broadband performance, offering deeper insights into potential improvements and the fundamental nature of a time-domain approach.

\subsection{Effects of the Unit Cell Dimensions and Spectral Weighting on Performance}
Motivated by observations from the previous section, we explore the impact of unit cell dimensions as optimization constraints on broadband performance. Additionally, we examine the impact of spectral weighting conveyed by the time signal during optimization.
\begin{figure*}[ht!]
\includegraphics[trim = 0mm 0mm 0mm 0mm,width=1\linewidth]{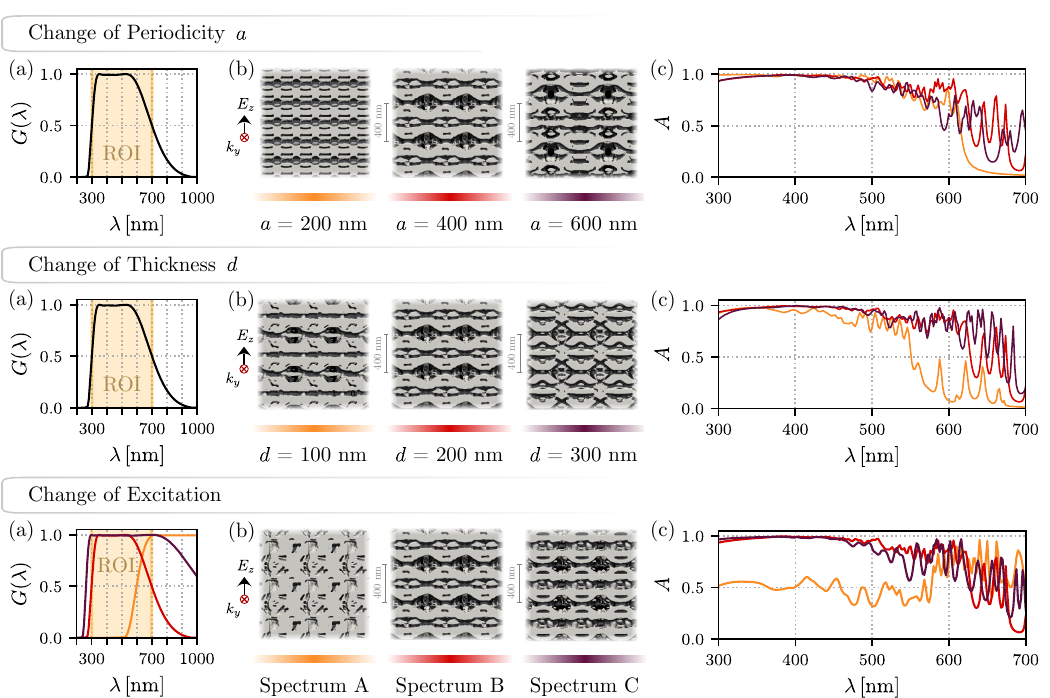}
\caption{\label{Fig:ParameterResults}\justifying Optimization results for different configurations of periodicity~$a$~(1st row), thickness~$d$~(2nd row) and spectral weighting~(3rd row) considered in the metasurface design. Each row represents a parametric sweep where one parameter varies while the other two remain fixed. The structure consistently shown across all sweeps (marked in red) corresponds to the original configuration from Fig.\,\ref{Fig:FirstResults}~(\mbox{$a=400$\,nm}, \mbox{$d=200$\,nm}) with the induced spectral weighting from Fig.\,\ref{Fig:Setup}(b), named 'Spectrum B' here. This design serves as the reference for the variations. (a)~Excitation spectrum $G(\lambda)$ conveyed by the time-domain excitation signal during optimization. The region of interest~(ROI)~\mbox{300–700 nm} to evaluate broadband performance is marked in yellow. (b)~Illustration of the optimized metasurfaces for each parametric variation, showing a section in the top-view. They all share the same length scale for comparability. (c)~Comparison of the absorption spectra within the ROI for the optimized metasurfaces under each variation.}
\end{figure*}

Fig.\,\ref{Fig:ParameterResults} presents the optimization results for the different hyperparameter settings, including periodicity~$a$ and thickness~$d$ and excitation spectrum. The spectral weighting is shown in (a), the designs in  top view in (b), and their absorption spectra in (c). The wavelength range~\mbox{300-700\,nm}, over which we aim to boost absorption, is hereafter called \textit{region of interest}~(ROI). The optimized metasurface and its configuration from the previous section serve as a reference for comparison across all parametric sweeps. Rapid variations in the absorption curves make a direct comparison of broadband performance challenging. We introduced a quantitative measure in Appendix~\ref{app:BroadBandQuant} to address this issue. In the following, we focus primarily on qualitative comparisons to observe performance trends as we vary the hyperparameters - with the quantitative measures summarized in Table~\ref{tab:ParameterResults}.

From the change in periodicity (Fig.\,\ref{Fig:ParameterResults}, 1st row), we observe an improved performance beyond \mbox{$\lambda>620$\,nm} for the periodicities \mbox{$400$\,nm} and \mbox{$600$\,nm} compared to \mbox{$200$\,nm}. In this range, characteristic sharp absorption spikes emerged, which we attribute to LRs. 
Both substrate ($n_{\text{SiO}_2}$) and superstrate (air) have their own RA at \mbox{$\lambda^{\text{RA}}_{\text{SiO}_2}=1.45\cdot a$} and  \mbox{$\lambda^{\text{RA}}_{\text{air}}=a$}. As the RA wavelengths redshift with increasing periodicity, the emergence of the spikes for both periodicities with \mbox{$a>200$\,nm} but absence for \mbox{$a=200$\,nm} can be explained. The LRs can only arise and be leveraged by the optimizer if the $\lambda^{\text{RA}}$s shift to the red regime, which was achieved as we increased the lattice constant. 
Across the ROI, no broadband improvement can be stated for a periodicity of \mbox{$600$\,nm} relative to \mbox{$400$\,nm}. As we noted an increase in our time-domain objective, however, we attribute this to the optimizer considering wavelengths across the entire excitation spectrum (see Fig.\,\ref{Fig:ParameterResults}(a), 1st row), including those beyond the ROI where the substrate’s Rayleigh anomaly occurs (at \mbox{$870$\,nm}). 

An improvement in broadband performance is clearly evident when increasing the thickness of the thin-film design layer (Fig.\,\ref{Fig:ParameterResults}, 2nd row). In fact, the design with \mbox{$d=300$\,nm} exhibits the best performance among all metasurfaces presented in this article. This tendency of improvement is expected. Initially, the optimizer is provided with more lossy material per unit cell to utilize. The out-of-plane extension of the unit cell increases the number of degrees of freedom for the evolving metasurface. Thus, it allows the formation of resonances at longer wavelengths and of higher-order localized and delocalized modes (with potential hybridization). We observed a denser clustering of absorption peaks in the red with increasing thickness, indicating a higher mode density of the LRs. 

Lastly, we study the dependence of performance on the spectral weighting under which the designs have evolved~(Fig.\,\ref{Fig:ParameterResults}, 3rd row). Each spectrum in~(a) belongs to a different spectral amplitude profile $G(\lambda)$, which was conveyed by the time signal of excitation during optimization. We labeled them by the letters `A', `B' and `C'. Spectrum~`B' is identical to that of Fig.\,\ref{Fig:Setup}(b), and was used in all optimization of the sweeps over periodicity and thickness. Spectrum~`C' provides a uniform weighting across the ROI and gives extra emphasis to wavelengths near the ROI's bounds relative to spectrum~`B'. Comparing both absorption spectra exhibited by the designs, we do not observe an overall improvement by this adjustment. This is not surprising, given our previous explanation that the optimization algorithm inherently prioritizes those wavelengths associated with greater material loss, i.e., primarily across the (ultra) violet. More interesting, however, is the evaluation of the performance of the design evolved under spectrum~`A'. Spectrum `A' spans a wavelength range that only partially overlaps with the ROI. Its lower (half-maximum)~bound is \mbox{$\lambda_{\text{1/2}}=600$\,nm}. In that case, Silicon's characteristic intrinsic losses cannot be leveraged during the optimization process. As we observe, the design shows a significantly reduced performance for wavelengths below~\mbox{$600$\,nm}, matching the spectral weight's lower bound~$\lambda_{\text{1/2}}$. Above that wavelength, its absorption is considerably higher and even exhibits absorption spikes that approach unity. This example demonstrates the effect of spectral weighting as a constraint. Initially, the optimizer can only utilize the material’s intrinsic (dispersive) response over the conveyed wavelengths. It relies on the exploitation of physical effects in that weighted spectrum to boost absorption most efficiently. For spectrum~`A', and given the initial (periodic) physical configuration, these are primarily the lattice-induced effects.


From this study on hyperparameters, we conclude: broadband performance can be improved through proper spectral weighting and by expanding the design’s degrees of freedom. 
However, dissipation is fundamentally bounded by both the design's dimension and the material’s intrinsic losses, limiting the extent to which absorption can be maximized \cite{Miller:16,lightTRLimit}. The achieved enhancement towards the red is accompanied by the emergence of narrow-band features and rapid spectral fluctuations associated with the LRs. 
This highlights the efficient exploitation of lattice effects by the time-domain adjoint method but also reveals the physical limitations of achieving uniform absorption.
\textcolor{black}{To corroborate our claim on the optimizer's ability to exploit the physics of collective resonances to boost absorptance at longer wavelengths, we also provide an additional study in Appendix~\ref{app:SLR_Opt}.}

\section{Conclusion} 
We enabled the inverse design of a superabsorbing semiconductor metasurface in the visible-to-ultraviolet spectral range. Starting from a thin-film silicon-on-glass configuration, we achieved this by (1) modeling silicon via CCPR and (2) employing our (FDTD) time-domain CCPR-ADE adjoint method to maximize the time-averaged power dissipation. We explored the physical effects on broadband performance that inherently arise from a metasurface’s periodic array configuration. It was observed that dimensional parameters and spectral weighting can be tailored to enhance absorption in towards red, where the improvement is primarily dictated by lattice-induced effects.
All designs evolved into all-connected metasurfaces, characterized by a complex interwoven network of waveguide-like structures. These topologies enabled high broadband absorption based on a complex interplay of physical phenomena. They motivate an in-depth analytical examination and can furthermore inspire conventional and other inverse design approaches; we therefore made these designs openly available~\cite{dataset}.

\textcolor{black}{In Appendix~\ref{app:Scaling}, we provide detailed information about the computational challenges of the employed time-domain adjoint algorithm using the FDTD method (in terms of time and memory complexity). Although the presented designs are not yet feasible with current lithographic methods, the TopOpt algorithm allows for a flexible incorporation of manufacturability constraints in addition~\cite{Hammond:21, Quantized, Probst:24}. In this work,} we chose Silicon as a representative semiconducting material, as it is the most widely used in the semiconductor industry. However, our CCPR-ADE formulation also allows for optimization of other dispersive media. These include metals, making it suitable for designing thermoplasmonic metasurfaces - or combining both dielectric and plasmonic materials for hybrid configurations.

Employing the time-domain adjoint method for nanophotonic inverse design is rarely reported in the literature, in contrast to its widely used frequency-domain counterpart. 
Our results highlight the power of the time-domain approach in achieving broadband performance, motivating further exploration and application to other photonic systems.

\appendices
\section{On a quantitative broadband measure}\label{app:BroadBandQuant}
We introduce here a simple quantitative measure of broadband performance that takes into account the sharp features and rapid fluctuations in the absorption spectra in Fig.\,\ref{Fig:ParameterResults}.

\begin{figure}[ht!]
  \begin{minipage}{0.48\textwidth}
    \centering
    \includegraphics[trim = 0mm 2mm 0mm 0mm, width=0.85\linewidth]{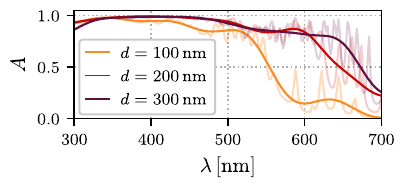}
    \captionof{figure}{\justifying Smoothed absorption spectra, using a Butterworth low-pass filter. Herein, applied to the spectra from Fig.\,\ref{Fig:ParameterResults}~(mid~row) for demonstration. Based on this filtering, we calculated the RMSD and $B_{\mathrm{F}}$ listed in Table~\ref{tab:ParameterResults}.}
    \label{Fig:LowPassFilter}
  \end{minipage}
  \hfill
  \vspace{0.2cm}
  \begin{minipage}{0.48\textwidth}
    \centering
    
    \begin{tabular}{|l||l|c|c|c|}
       \hline
      & \makecell{Parameters\\($a$, $d$, Exc.)}
      & $\langle A\rangle$
      & $B_{\mathrm{F}}$
      & $\mathrm{RMSD}$ \\
      \hline
      Reference  & (400\,nm, 200\,nm, B) & 0.824 & 0.732 & 0.085 \\
      \hline
      \multirow{2}{*}{Periodicity}
        & (200\,nm, 200\,nm, B) & 0.737 & 0.684 & 0.045  \\
        & (600\,nm, 200\,nm, B) & 0.793 & 0.715 & 0.074 \\
      \hline
      \multirow{2}{*}{Thickness}
        & (400\,nm, 100\,nm, B) & 0.625 & 0.595 & 0.063  \\
        & (400\,nm, 300\,nm, B) & 0.856 & 0.768 & 0.083  \\
      \hline
      \multirow{2}{*}{Excitation}
        & (400\,nm, 200\,nm, A) & 0.549 & 0.582 & 0.084  \\
        & (400\,nm, 200\,nm, C) & 0.823 & 0.755 & 0.091  \\
      \hline
    \end{tabular}
    \captionof{table}{ \justifying Average~$\langle A \rangle$, fractional bandwidth~$B_{\mathrm{F}}$, and the root mean square deviation~(RMSD) of the absorption spectra from Fig.\,\ref{Fig:ParameterResults}. $B_{\mathrm{F}}$ and RMDS were computed with respect to a low-pass filtered signal~(Fig.\,\ref{Fig:LowPassFilter}).}
    \label{tab:ParameterResults}
  \end{minipage}
\end{figure}

The simplest way to quantify enhanced absorption is to compute the average $\langle A \rangle$, evaluated over the wavelength range of interest. However, this metric alone does not capture spectral variations, making it unable to distinguish between a uniformly broadband response and a narrow peak. To address this limitation, we additionally evaluate the fractional bandwidth $B_{\mathrm{F}}$ and the root mean square deviation (RMSD) with respect to a smoothed signal. To suppress sharp oscillations in the absorption spectra, we apply a low-pass filter on the spectra. Specifically, we choose a 4th-order Butterworth filter with (normalized) cutoff frequency 0.05~\cite{Butterworth1930}~(using \textit{Scipy’s} \texttt{butter}), as illustrated in Fig.\,\ref{Fig:LowPassFilter}. This filtering ensures a stable measure of~$B_{\mathrm{F}}$ and quantification of variations in the (original) absorption spectrum. The RMSD provides a statistical measure of fluctuations across the entire wavelength range, while the fractional bandwidth~$B_{\mathrm{F}}$ characterizes the relative spectral range where absorption remains significant~($A > 0.5$). Together with the average, these three parameters establish a systematic quantitative evaluation of broadband performance.

Table~\ref{tab:ParameterResults} summarizes these quantities for the absorption spectra presented in Fig.\,\ref{Fig:ParameterResults}. From that, we can state that the design with \mbox{$d=300$\,nm} exhibits the overall best broadband performance, as indicated by the largest values of both $\langle A \rangle$ and $B_{\mathrm{F}}$. In contrast, the structure evolved under the excitation spectrum~`A' performs the worst. A relatively high RMSD they both have in common is caused by the rapid variations in the red regime, which we attributed to lattice resonances.

\section{{\color{black}Enhancing the absorption from a lattice resonance via time-domain TopOpt}}\label{app:SLR_Opt}

To further support our statement in Sec.~\ref{Sec:Opt} on the ability of the optimizer to efficiently exploit surface lattice resonances~(SLRs) to enhance absorption at longer wavelengths, we provide an additional study here. We chose a configuration with disconnected meta-atoms, which allows for analytical identification of the metasurface’s collective resonances using multipole decomposition techniques and the coupled-dipole model~\cite{Allayarov2023}. This study also highlights a strategy for time-domain optimization over a narrow spectral range -- a problem whose efficient tackling is usually reserved for frequency-based methods.

\subsection{Optimization setup\label{Sec:optset}}
\begin{figure}[ht!]
\includegraphics[width=1\linewidth]{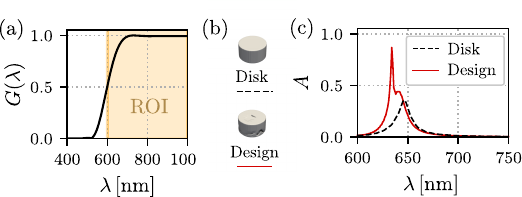}
\caption{\label{Fig:CylinderResults} (a) Spectral weighting of the time-domain signal delivered by the pulse that excites the evolving design at each TopOpt iteration. The spectrum corresponds to that of `Spectrum~A' in Fig.6~(3rd row) in Section~\ref{Sec:Opt}. (b) Silicon Nanodisk and (constrained) TopOpt design in a single unit cell of their respective periodic array configuration, and (c) their absorption over broadband as a response to a plane wave at normal incidence.}
\end{figure}

We consider an (infinite) periodic array configuration of Silicon particles in a homogeneous environment~\mbox{($n_{\text{SiO$_{\text{2}}$}}=1.46$)}, i.e., the permittivities of substrate and superstrate are equal. For a configuration with periodicity~\mbox{$a=400$\,nm} of nanodisks with radius~\mbox{$R=90$\,nm} and height \mbox{$h=100$\,nm} (excited by a plane wave at normal incidence), we observed an isolated absorption response in the red (around \mbox{$\lambda=646$\,nm}); we associated this feature with an SLR. 
For the optimization, we start from that nanodisk as the initial design. By restricting the design's expansion to these cylindrical bounds, we let the design evolve under a spectral weighting that has a lower (half-maximum)~bound~\mbox{$\lambda_{\text{1/2}}=600$\,nm} and prioritizes wavelengths in the (infra) red (see Fig.\,\ref{Fig:CylinderResults}(a)).
This approach is motivated by the following idea: as reported in Ref.~\cite{Castellanos2019}, the SLR redshifts as the cylinder's diameter increases (keeping the periodicity fixed). We aim to enhance absorption related to a single SLR by (1)~constraining the design's spatial expansion to prevent the resonance's redshift while the structure evolves, and (2)~by suppressing the optimizer's exploitation of optical responses contributing to the dissipation for smaller wavelengths \mbox{$\lambda<\lambda_{\text{1/2}}=600$\, nm}, as the excitation spectrum does not cover them (or their spectral weight is only small).

\begin{figure}[ht!]
\includegraphics[width=1\linewidth]{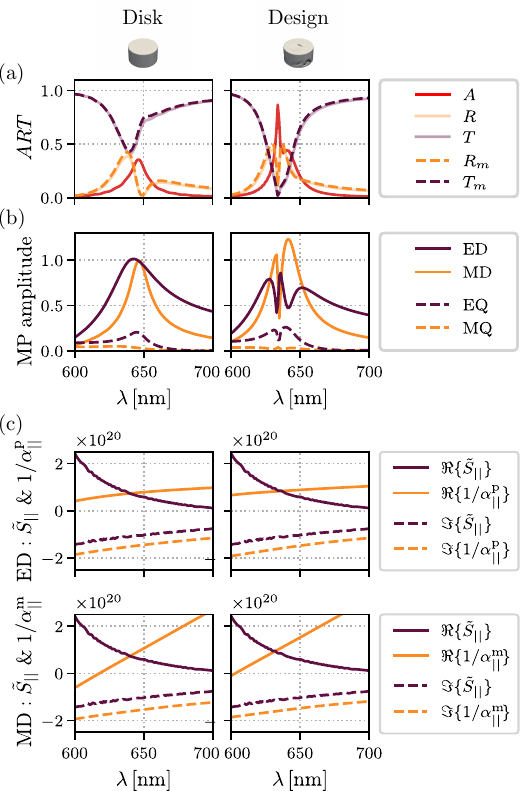}
\caption{\label{Fig:Analysis}(a)~Absorptance~(A), transmittance~(T), and reflectance~(R) of the periodic regular (left) and optimized (right) disks presented in Fig.\,\ref{Fig:CylinderResults}. The solid (dashed) curves correspond to the full numerical (multipole approach) calculations. (b)~Amplitudes of multipole~(MP) contributions in the reflection and transmission coefficients, where ED~(MD) is the electric~(magnetic) dipole, and EQ~(MQ) is the electric~(magnetic) quadrupole. (c)~Visualization of the SLR conditions~\mbox{(\ref{Eq:SLRCondED})\,\&\,(\ref{Eq:SLRCondMD})} for electric (top sub-panel) and magnetic (bottom sub-panel) dipole; $1 / {\alpha}_{||}$ and $\tilde{S}_{\parallel}$ have the units of m$^{3}$. 
}
\end{figure}

Fig.\,\ref{Fig:CylinderResults}(b) shows the initial nanodisk and the design that has evolved under constraints to the cylindrical bounds. Their absorption spectra up to \mbox{$750$\,nm} are shown in Fig.\,\ref{Fig:CylinderResults}(c); beyond that wavelength, no significant variation of the absorption was observed anymore and is essentially zero. We identify the spectrally isolated absorption response of the nanodisk with a maximum of \mbox{$A_{\text{max}}=0.356$} at \mbox{$\lambda=646$\,nm}. The spectrum of the restricted design shows a similar (enhanced) absorption profile as the original nanodisks, but with an additional emerged narrow feature at \mbox{$\lambda=634$\,nm} with \mbox{$A_{\text{max}}=0.868$}. It is closely neighbored in the spectrum to that of the original nanodisks, but slightly shifted towards smaller wavelengths. The periodic configuration of both the nanodisk and TopOpt design in a homogenous environment allows for employing the coupled-dipole model to identify SLR analytically~\cite{Allayarov2023}, which will be outlined below. 

\subsection{SLR condition in dipole approximation} 

For a given periodic square array of disks located in a homogeneous environment and irradiated by a $z$-polarized plane wave propagating along the $y$-axis, the electric dipole moment of each disk is~\cite{garcia2007colloquium,Allayarov2023}
\begin{equation}
p_z=\frac{\varepsilon_0 \varepsilon_{\mathrm{d}} {E}_z}{1 /{\alpha}_{\parallel}^{\mathrm{p}}-\tilde{S}_{\parallel}},
\end{equation}
where $\varepsilon_0$ and $\varepsilon_d$ are the vacuum permittivity and the superstrate's relative permittivity, respectively, ${E}_z$ is the incident electric field at the geometrical disk center (in the absence of the disk), ${\alpha}_{\parallel}^{\mathrm{p}}$ is the in-plane electric dipole polarizability of a single disk and ${\tilde{S}}_{\parallel}$ is the in-plane component of the dipole lattice sum associated with the electromagnetic interaction (coupling) between electric dipoles located at the lattice nodes of the metasurface. 

From the above equation, one observes that an SLR associated with the electric dipole~(indicated by the superscript ``$\mathrm{p}$'') emerges at a wavelength for which
\begin{equation}\label{Eq:SLRCondED}
\Re\left\{1 / {\alpha}_{\parallel}^{\mathrm{p}}\right\}=\Re\left\{{\tilde{S}}_{\parallel}\right\},
\end{equation}
and the quality factor of the resonance scales as \mbox{$Q_{\text{f}}\propto 1/\Im\{1 / {\alpha}_{\parallel}^{\mathrm{p}}-{\tilde{S}}_{\parallel}\}$}.  A similar condition for the SLR associated with the magnetic dipole~(indicated by the superscript ``$\mathrm{m}$'') is given by
\begin{equation}\label{Eq:SLRCondMD}
\Re\left\{1 / {\alpha}_{\parallel}^{\mathrm{m}}\right\}=\Re\left\{{\tilde{S}}_{\parallel}\right\}.
\end{equation}

\subsection{SLRs of periodic regular and optimized disks}

We conducted a multipole analysis and the exploration of the SLR for both periodic configurations presented in Fig.\,\ref{Fig:CylinderResults}. First, we analyze the spectrum of the periodic regular disks (left column). From Fig.\,\ref{Fig:Analysis}(a), we see that the structure has a resonance around $\lambda=640$~nm. As one can observe from Fig.\,\ref{Fig:Analysis}(b), this resonance occurs primarily due to the lattice electric and magnetic dipole resonances (SLRs). Indeed, by plotting the SLR conditions~\mbox{(\ref{Eq:SLRCondED})\,\&\,(\ref{Eq:SLRCondMD})} in Fig.\,\ref{Fig:Analysis}(c), we see that both electric and magnetic SLRs are predicted to appear at that wavelength. The dip in the reflectance occurring near $\lambda=650$~nm is due to the destructive interference between electric and magnetic dipole resonance, also known as the lattice Kerker~\cite{babicheva2017resonant}. The dip in the transmittance occurring around $\lambda=640$~nm can be interpreted as a partial lattice anti-Kerker effect, since the required amplitude and phase conditions for the electric and magnetic dipoles are not fully realized.

\textcolor{black}{We evaluate the optical response of the optimized disks, as shown in the right column of Fig.\,\ref{Fig:Analysis}.} One can see that the main difference from that of the periodic regular disks is the existence of a narrow dip in the reflectance and transmittance at $\lambda=634$~nm~(Fig.\,\ref{Fig:Analysis}(a)). The multipole analysis of this resonance in Fig.\,\ref{Fig:Analysis}(b) shows that both electric and magnetic dipoles (and weak quadrupoles as well) have a narrow resonance around that particular wavelength, too. The SLR conditions~~\mbox{(\ref{Eq:SLRCondED})\,\&\,(\ref{Eq:SLRCondMD})} predict a collective lattice resonance around that wavelength as well~(Fig.\,\ref{Fig:Analysis}(c)), from which one can attribute the peaks in the spectrum to SLRs. The associated absorptive response has a very prominent peak; this is due to the fact that the quality factor $Q_{\text{f}}$ of the dipole resonances at that wavelength is relatively high, and hence, they have a longer lifetime to get absorbed~\cite[App.~D]{GedeonDissip}.  
\textcolor{black}{This result can also be interpreted as an attempt of the optimizer to bring both transmittance and reflectance to zero by realizing the above-mentioned lattice Kerker and anti-Kerker effects nearly at the same wavelength, also known as balanced Kerker effect~\cite{alaee2017theory}.
This effect contributed to an enhanced dissipation that drove the gradient-based optimization algorithm and thus demonstrates its efficiency in the exploitation of lattice resonances.}

\section{{\color{black}Time and memory complexity of the time-domain adjoint algorithm using the FDTD~method}}\label{app:Scaling}
Here, we provide an overview of the computational cost and scaling of our time‑domain adjoint method used in a (3D)~FDTD~framework. The prerequisites for the explanation can be found in the Supplementary Material~(SM), Sec.~I-III, to which we will refer here. 

FDTD~forward and adjoint simulation, stand-alone, are inherently bounded in modelling arbitrarily large-scale problems (in terms of \textit{space and time}) due to the constraints on the step sizes~$\{\Delta x, \Delta y, \Delta z\}$ and $\Delta t$ inherent to the FDTD method~\cite{Taflove}. The simulation of a nanostructure requires selecting a spatial step size small enough to both (1) capture the structure’s geometry accurately and (2) minimize the effects of numerical dispersion (the artificial distortion of a wave as it propagates through the FDTD computational grid \cite[Ch. 2.6]{Taflove}). Moreover, spatial and temporal step sizes are not independent from each other, governed by the  \textit{Courant-Friedrichs-Lewy}~(CFL)~condition; it sets an upper bound of the time step size~\cite[Ch. 4.7.1]{Taflove}~\footnote{\label{fn:criteria}%
  The criterion expression presented here serves only as a \emph{rule of thumb},
  as it is derived from the assumption that the wave propagates through a
  homogeneous, lossless (non-magnetic) dielectric. In general, the required fineness of the resolutions (and CFL~condition) depends on the physical problem considered, and a sufficient accuracy can be more reliably guaranteed by comparison with analytical results (e.g. from Mie analysis)~\cite{Lesina:15, GedeonDissip} or experimental measurements~\cite{Pala2016}.}
\begin{align}
  \Delta t 
    \le \frac{1}{c \sqrt{\frac{1}{\Delta x^2}
                       + \frac{1}{\Delta y^2}                   + \frac{1}{\Delta z^2}}},
      \label{Eq:CFL}
\end{align}
where $c$ denotes the (maximum)~speed of light in the system. This inequality ensures the stability of the FDTD~algorithm by preventing the physical wave from propagating faster than the information can even proceed through the numerical grid.\newline
Approximating subwavelength geometrical features together with reducing the effects of numerical dispersion restricts an arbitrary choice of the physical domain's expansion, as a sufficiently small spatial step size is required to capture the effects of a nanostructure accurately. Due to the CFL condition~(\ref{Eq:CFL}), a higher spatial resolution also implies a higher resolution in time. Considering a fixed physical time period~$T$, this results in a larger number of field values that need to be computed. Since the discretized fields in time~(SM Sec.~I.A) need to be updated in each Yee cell~(SM Fig.\,S2), the \textit{time complexity} for a single FDTD~run scales with $O(NM)$, where ``$N$'' denotes number of \textit{Yee cells} in the FDTD~simulation domain and ``$M$''~denotes the number of time steps. In 3D, that implies that a doubling of the spatial extent along each dimension $x$, $y$ and $z$ would yield an increase of simulation time by a factor~$2^{3}=8$.

A larger number of time steps in the FDTD simulations also leads to higher memory demands when used in the adjoint method, since the electric (and auxiliary)~fields must be stored in both space and time to compute the gradient.
If ``$N_{\Omega}$'' denotes the number of Yee cells in the design region and ``$P$''~the number of the design material's CCPR~poles~(SM Eq.~(S1)), one observes from the discretized gradient expression~(SM Eq.\,(S27)) that the \textit{memory complexity} scales at least with $O(N_{\Omega}\, M + P\, N_{\Omega} \, M)$ (assuming a non-dispersive background medium, and not considering the storage related to any specified objective defined on $\Omega_{\mathrm{o}}$). We note that the number of design variables is in fact~$3\times N_{\Omega}$~(SM Sec.~III.A), due to their assignment to three electric field components within the staggered 3D~Yee~grid~(SM Fig.\,S2). The first term within the \mbox{big~$O$~expression} refers to memory demands for storing the (real-valued)~electric field, while the second term refers to those of the (real part) of the complex auxiliary fields. This expression provides an intuitive understanding of the high memory requirements inherent to the time-domain adjoint method. It reveals the significant increase in memory if dispersion is considered at all~($P>0$). The term~``$N_{\Omega}\, M$" further implies a restriction on optimization of large design structures and their observation over a long time period~$T$. The latter can be problematic for special configurations that might be considered. For example, a system that exhibits a high-\textit{Q} resonance associated with a long-lasting decay of the fields over time. This might require a premature termination of the simulations and can therefore hinder the convergence to solutions using gradient-based solvers, as the gradients are not captured accurately. A further issue is posed by the attempt to maximize this sharp resonance directly. As it represents a narrow response in the frequency domain, the tailored time signal injected into the system must be quasi-infinite~($T\rightarrow\infty$) to maximize exclusively the single-frequency response. However, under certain conditions and with a tailored strategy, it is possible to optimize efficiently over a narrow band (as we demonstrated in Appendix~\ref{app:SLR_Opt}).

In the optimizations presented in this article, the absorption is merely dictated by the lossy optimized structure surrounded by \textit{lossless} dielectrics. The FDTD adjoint method allows flexibility to tackle other configurations than those presented, e.g., considering more complex setups such as hybrid plasmonic-dielectric systems, a lossy background medium, or (most generally) multiple dispersive, lossy media. As all lossy materials contribute to the absorption mechanism, one needs to account for the storage of the fields and the inclusion in the adjoint algorithm of all involved media to enhance the total absorption of the (entire) physical system most efficiently; the optimization is driven by the maximization of the local power dissipation~(SM Sec.~II) that would be affected by a complex interplay of the different materials' (sub)structures.

For our optimizations, we employed a high-performance computing approach using FDTD~domain decomposition~\cite{parallelism}, which enables both speed-up of the algorithm and solves the issue with the memory demands by an efficient workload~distribution across multiple computer nodes equipped with CPUs. 

\bibliographystyle{IEEEtran.bst}
\bibliography{main.bib}
\end{document}


\title{Supplementary Material for \\
``Topology optimization of a superabsorbing thin-film semiconductor metasurface''}
\author{Johannes Gedeon, Izzatjon Allayarov, Emadeldeen Hassan, and Antonio Cal{\`a} Lesina
}

\maketitle
This supplementary material provides an introduction to time-domain Maxwell's equations using auxiliary differential equations~(ADEs) based on the \textit{Complex-Conjugate Pole-Residue Pairs}~(CCPR)~model to simulate dispersive media in the \textit{Finite-Difference Time-Domain Method}~(FDTD) (Sec.~\ref{Sec:CCPR_ADEs}). For the reader's convenience of Eq.~(2) of the main article, we derive the expression for the instantaneous electric power dissipation density in Section~\ref{sec:derivations_powerdissip}, and present the key results of derivations of the time-domain adjoint algorithm for density-based topology optimization~(TopOpt) in Section~\ref{Sec:AdjointMethod} (reproduced from our previous work~\cite{GedeonDissip}). 

In addition to the electric field and dissipation profiles in Fig.~4\,(c) and Fig.~5 of the main article, an illustration of the power flow through the Ridges structure is given in Section~\ref{Sec:PowerFlow} for further physical insight.


\section{CCPR-ADE formulation of the time-domain Maxwell's equations for the FDTD method.}\label{Sec:CCPR_ADEs}
\subsection{The CCPR~Model}\label{sec:CCPR_model}
\begin{figure}[h]
\centering
\includegraphics[width=0.9\linewidth]{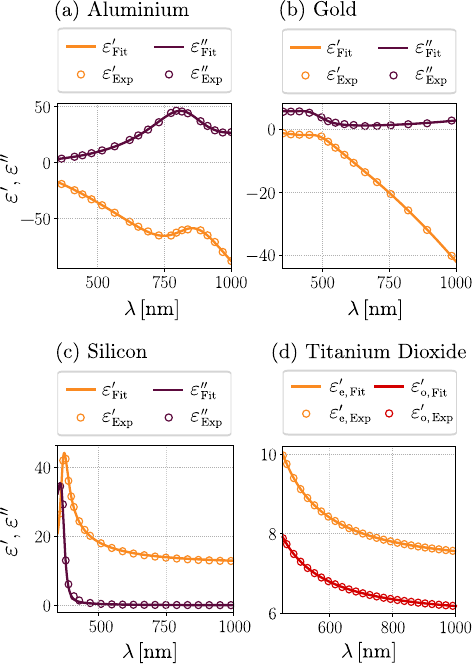}
\caption[Modelling of the complex relative permittivity~$\varepsilon'-j\varepsilon''$ of Aluminium, Gold, Silicon, and (anisotropic)~Titanium dioxide by using the CCPR~model]{Modelling of the complex relative permittivity~$\varepsilon'-j\varepsilon''$ of (a)~Aluminium, (b)~Gold, (c)~Silicon, and (d)~Titanium dioxide by using the CCPR~model~(\ref{Eq:CCPR_isotropic}) in the wavelength range~\mbox{350-1000 nm}; the experimental data used are reported in the Refs.~\cite{Rakic:95, Au_Johnson, Si_Schinke, DeVore:51}, respectively. The permittivity of the anisotropic Titanium dioxide consists of an ordinary~(``o'') and extraordinary~(``e'') part, which were fitted separately. The corresponding CCPR~parameters can be found in Ref.~\cite{Gedeon2023}.}
\label{CCPR_Plot} 
\end{figure}

The CCPR~model allows for an accurate description of the frequency dispersion of a linear medium over broad spectral ranges~\cite{CCPRmodel}. It can be seen as a generalization of the conventional Drude and Lorentz models, and thus provides flexibility in modelling the dielectric response of arbitrary media over broadband when incorporated into the \textit{Finite-Difference Time-Domain}~(FDTD) method~\cite{Taflove} via auxiliary differential equations~(which will be explained in Sec.~\ref{Sec:ADEs}). Considering a $\mathrm{e}^{\,j\omega t}$~time-dependence and isotropic media, the model reads
\begin{equation}\label{Eq:CCPR_isotropic}
\varepsilon(\omega)=\varepsilon_{\infty}+\frac{\sigma}{j \omega \varepsilon_0}+\sum_{p=1}^{P}\left(\frac{c_{p}}{j \omega-a_{p}}+\frac{c_{p}^*}{j \omega-a_{p}^*}\right),
\end{equation}
where $\varepsilon_{\infty}$ is the relative permittivity at infinite frequency, and  $\sigma$ is the static conductivity. 
The parameters~$a_p$ and $c_p$ appearing in the pole terms are complex and ``$*$'' denotes their complex conjugation. Fig.~\ref{CCPR_Plot} shows the CCPR-fits of the permittivity for different metallic and dielectric materials in the wavelength range~\mbox{350-1000 nm}.  

\subsection{CCPR-ADEs}\label{Sec:ADEs}
A well-established way to incorporate a dispersion model into the FDTD method is the use of auxiliary differential equations~(ADEs), which can be updated in synchronism with Maxwell’s curl equations~\cite{HanCCPRADEs}. For simplicity, we will hereafter limit our explanation to the derivation of the CCPR-ADEs in their \textit{analytical} form for non-magnetic, isotropic media. The following steps are taken from Ref.~\cite{CCPRmodel}, which also covers the general case of anisotropic dispersion and provides a FDTD update scheme in time. 

The \textit{time-harmonic} Maxwell’s equations in a linear and source-free medium are taken as the starting point. For a (non-magnetic) dispersive medium with $\mathrm{e}^{\,j\omega t}$~dependence, the equations read
\begin{subequations}\label{Eq:Maxwell_freq}
\begin{align}
\nabla \times \hat{\mathbf{E}} & =-j \omega \mu_0 \hat{\mathbf{H}}, \\
\nabla \times \hat{\mathbf{H}} & =\phantom{-}j \omega \varepsilon_0 \varepsilon(\omega) \hat{\mathbf{E}},
\end{align}
\end{subequations}
where $\varepsilon$ is the relative permittivity. 

Inserting the CCPR~expression of $\varepsilon(\omega)$~(\ref{Eq:CCPR_isotropic}) into Eq.~(\ref{Eq:Maxwell_freq}b) yields
\begin{equation}\label{Eq:curlH_freq}
(\nabla \times \hat{\mathbf{H}})= j \omega \varepsilon_0 \varepsilon_{\infty} \hat{\mathbf{E}} + \sigma \hat{\mathbf{E}} +j \omega \sum_{p=1}^{P}\left(\hat{\mathbf{Q}}_{p}+\hat{\mathbf{Q}}_{p}^{\prime}\right),
\end{equation}
with the auxiliary fields
\begin{subequations}\label{Eq:AuxFields_freq}
\begin{align}
\hat{\mathbf{Q}}_{p}&=\frac{\varepsilon_0 c_{p}}{j \omega-a_{p}} \hat{\mathbf{E}}, \\
\hat{\mathbf{Q}}_{p}^{\prime}&=\frac{\varepsilon_0 c_{p}^{*}}{j \omega-a_{p}^{*}} \hat{\mathbf{E}}.
\end{align}
\end{subequations}
By multiplying these equations by their denominators and subsequently applying the Fourier transformation, one obtains the time-domain equations of the auxiliary fields as
\begin{subequations}\label{Eq:AuxFields_timedep}
\begin{align}
& \partial_t\mathbf{Q}_{p}-a_{p} \mathbf{Q}_{p}=c_{p} \varepsilon_0 \mathbf{E}, \\
& \partial_t\mathbf{Q}_{p}^{\prime}-a_{p}^* \mathbf{Q}_{p}^{\prime}=c_{p}^{*} \varepsilon_0 \mathbf{E}.
\end{align}
\end{subequations}
As the electric field in the time domain is considered to be real, one observes from the complex conjugation of Eq.~(\ref{Eq:AuxFields_timedep}a) that $\mathbf{Q}^{*}_{p}\stackrel{!}{=}\mathbf{Q}_{p}^{\prime}$. By using this relation, and transforming Eqs.~(\ref{Eq:curlH_freq}) and (\ref{Eq:Maxwell_freq}a) into the time domain, the Maxwell's equations in time read
\begin{subequations}\label{Eq:Maxwells2}
  \begin{empheq}[]{align}
-(\nabla \times \mathbf{H}) + \varepsilon_{0}\varepsilon_{\infty}\partial_{t}\mathbf{E}+ \sigma \mathbf{E} + 2\sum_{p=1}^{P}\Re\left\{\partial_t\mathbf{Q}_{p}\right\} &= 0,\\[1pt]
\forall p \in \{1, \ldots, P\}: \partial_{t}\mathbf{Q}_{p}-a_{p}\mathbf{Q}_{p}- \varepsilon_{0}c_{p}\mathbf{E} &= 0,\\[8pt]
\mu_0 \partial_t \mathbf{H}+\nabla\times\mathbf{E} &=0.
\end{empheq}
\end{subequations}
This system of equations accounts for the electric response of dispersive media over the entire spectral range of frequencies conveyed by the time signal of an incident pulse. It can be discretized in space and time to provide an iterative update scheme for the FDTD~method.

\subsection{The FDTD update equations in time}\label{Sec:CCPR_updateEqs}
Here, we present the (time) discretization of Maxwell's equations using CCPR-ADEs~(\ref{Eq:Maxwells2}) that provides an update scheme for the FDTD method. The steps presented in the following are taken from Ref.~\cite{CCPRmodel}.

The auxiliary equations~(\ref{Eq:Maxwells2}b) can be discretized at time steps~$(m + 1/2)\Delta t$ by employing a first-order finite difference approximation of~$\partial_{t}\mathbf{Q}_{p}$ and the arithmetic mean of~$\mathbf{Q}_{p}$ and $\mathbf{E}$, respectively,
\begin{equation}\label{Eq:Q_update}
\mathbf{Q}_{p}^{m+1}=\frac{2+a_{p} \Delta t}{2-a_{p} \Delta t} \mathbf{Q}_{p}^m+\frac{\varepsilon_0 c_{p} \Delta t}{2-a_{p} \Delta t}\left(\mathbf{E}^{m+1}+\mathbf{E}^m\right).
\end{equation}
Discretizing the curl equation~(\ref{Eq:Maxwells2}a) at time steps~$(m+1/2)\Delta t$ as well, and inserting the update expressions for the auxiliary values from above, yields an equation of the form
\begin{equation}\label{Eq:Update_SystemLinearEquations}
\alpha \, \mathbf{E}^{m+1}=\mathbf{G}^m.
\end{equation}
The constant~$\alpha$ depends on the CCPR~model's parameters only,
\begin{equation}\label{Eq:E_update}
\alpha = \frac{\varepsilon_0 \varepsilon_{\infty}}{\Delta t} + \frac{\sigma}{2} + \frac{2}{\Delta t} \sum_{p=1}^{P} \Re \left\{ \frac{\varepsilon_0 c_{p} \Delta t}{2 - a_{p} \Delta t} \right\}.
\end{equation}
The vector~$\mathbf{G}^m$ is updated every time step according to
\begin{align}
    \mathbf{G}^m &= \left( \nabla \times \mathbf{H} \right)^{m+1/2} \\ \notag
&+ \left(\frac{\varepsilon_0 \varepsilon_{\infty}}{\Delta t} - \frac{\sigma}{2} - \frac{2}{\Delta t} \sum_{p=1}^{P} \Re \left\{ \frac{\varepsilon_0 c_{p} \Delta t}{2 - a_{p} \Delta t} \right\}\right)  \mathbf{E}^m \\ \notag
&- \frac{2}{\Delta t} \sum_{p=1}^{P} \Re \left\{ \frac{2 a_{p} \Delta t}{2 - a_{p} \Delta t} \mathbf{Q}_{p}^m \right\}. 
\end{align}
As it can be seen from Eq.~(\ref{Eq:Update_SystemLinearEquations}), computing $\alpha$ and $\mathbf{G}^m$ provides the update equation of the electric field for time steps~$(m + 1)\Delta t$. The update equation for the magnetic field~$\mathbf{H}$ in Eq.~(\ref{Eq:Maxwells2}c) takes the simple form 
\begin{equation}\label{Eq:H_update}
\mathbf{H}^{m+\frac{1}{2}} = \mathbf{H}^{m-\frac{1}{2}} - \frac{\Delta t}{\mu_0} (\nabla \times \mathbf{E}^m).    
\end{equation}
The update procedure for each time iteration can be summarized as follows:
\begin{enumerate}
    \item Update $\mathbf{H}$ according to Eq.~(\ref{Eq:H_update});
    \item Store the current values of $\mathbf{E}$;
    \item Update $\mathbf{E}$ according to Eq.~(\ref{Eq:Update_SystemLinearEquations});
    \item \mbox{$\forall p \in \{1, \ldots, P\}:$} Update the auxiliary field $\mathbf{Q}_{p}$ according to Eq.~(\ref{Eq:Q_update}).
\end{enumerate}

\section{The expression of the instantaneous electric power dissipation density based on CCPR-ADEs}\label{sec:derivations_powerdissip} 
Here, we provide the derivation of the instantaneous electric power dissipation density based on the CCPR-ADEs formulation of Maxwell's equations in the time domain~(\ref{Eq:AuxFields_timedep}). A procedure similar to that in Ref.~\cite{Shin:12} is followed, which is therein restricted to materials described by the Lorentz model. 

The Poynting's theorem in the time domain for (generally) dispersive, lossy media reads
\begin{equation}\label{Eq:PoyntingTheorem_1}
-\oint_{\partial\Omega_{\mathrm{o}}}(\mathbf{E} \times \mathbf{H}) \cdot \mathrm{d} \mathbf{a}=\int_{\Omega_{\mathrm{o}}}\left(\mathbf{E} \cdot \partial_t\mathbf{D}+\mathbf{H} \cdot \partial_t\mathbf{B}\right) \mathrm{d}^3 r.
\end{equation}
Using the ansatz,
\begin{align}
\mathbf{E} \cdot \partial_t\mathbf{D}:=& \;\partial_t u_e\, + q_e,\label{Eq:AnsatzE}\\
\mathbf{H} \cdot \partial_t\mathbf{B}:=& \;\partial_t u_m + q_m,
\end{align}
the theorem can be reformulated as
\begin{equation}\label{Eq:PoyntingTheorem_2}
\begin{split}
-\oint_{\partial\Omega_{\mathrm{o}}}(\mathbf{E} \times \mathbf{H}) \cdot \mathrm{d} \mathbf{a}=&\phantom{+}\int_{\Omega_{\mathrm{o}}}\left(\partial_t u_e+\partial_t u_m\right) \mathrm{d}^3 r \\
&+\int_{\Omega_{\mathrm{o}}}\left(q_e+q_m\right)  \mathrm{d}^3 r.
\end{split}
\end{equation}
Here, $q_e(t)$ and $q_m(t)$ are the instantaneous electric and magnetic \textit{power dissipation} densities, and $u_e(t)$ and $u_m(t)$ are the instantaneous electric and magnetic \textit{energy} densities, respectively. The curl expression of the magnetic field within the  CCPR-ADE~formulation~(\ref{Eq:Maxwells2}a) reads
\begin{equation}
\nabla \times \mathbf{H}=\varepsilon_0 \varepsilon_{\infty} \partial_t \mathbf{E}+\sigma \mathbf{E}+2 \sum_{p=1}^{P} \Re\left\{\partial_t \mathbf{Q}_{p}\right\}.
\end{equation}
Using the relation between electric and auxiliary fields from Eq.~(\ref{Eq:Maxwells2}b), assuming~$c_p \neq 0$,~$\forall p \in \{1, \ldots, P\}$,
\begin{equation}
\mathbf{E} = \frac{\partial_t \mathbf{Q}_{p}}{\varepsilon_{0}c_p} - \frac{a_p\mathbf{Q}_p}{\varepsilon_{0}c_p},
\end{equation}
and by recalling the Ampère-Maxwell~law $\nabla \times \mathbf{H} = \partial_{t} \mathbf{D}$, the product~$\mathbf{E}\cdot\partial_t \mathbf{D}$ can be expressed as
\begin{equation}\label{Eq:Dens+Dissip_2ndForm}
\begin{split}
\mathbf{E}\cdot\partial_t \mathbf{D}=&\phantom{+}\partial_{t}\left(\dfrac{1}{2}\varepsilon_0 \varepsilon_{\infty} \mathbf{E}^{2}-\sum_{p=1}^{P} \Re\left\{\frac{a_p\mathbf{Q}_{p}^2}{\varepsilon_{0}c_p}\right\}\right) \\
&+\sigma \mathbf{E}^{2}+2 \sum_{p=1}^{P} \Re\left\{\frac{(\partial_{t}\mathbf{Q}_p)^2}{\varepsilon_{0}c_p}\right\}.
\end{split}
\end{equation}
Identifying $\mathbf{E}\cdot\partial_t \mathbf{D}={\color{black}\partial_{t}} u_e + q_e$ from Eq.~(\ref{Eq:AnsatzE}), the instantaneous electric power dissipation density must be
\begin{equation}\label{Eq:InstantanousDissipation}
q_e(t) = \sigma \mathbf{E}^{2} + 2\sum_{p=1}^{P} \Re\left\{\frac{(\partial_{t}\mathbf{Q}_p)^2}{\varepsilon_{0}c_p}\right\}.
\end{equation}.

To measure the broadband performance of an optimized device in the \textit{steady state}, we consider the time harmonic fields $\mathbf{E}(t):= \Re\{\hat{\mathbf{E}}e^{j\omega t}\}$, $\mathbf{H}(t):= \Re\{\hat{\mathbf{H}}e^{j\omega t}\}$, oscillating with a frequency $\omega$; the amplitudes $\hat{\mathbf{E}}$, $\hat{\mathbf{H}}$ are frequency independent. Averaging Poynting's theorem~(\ref{Eq:PoyntingTheorem_2}) over an optical cycle yields~\cite[Ch. 6.9]{jackson} \cite[\S 61]{landau1995electrodynamics} \cite[App. A]{GedeonDissip},
\begin{equation}\label{Eq:PoyntingVector}
\nabla\cdot\langle\mathbf{S}\rangle=\nabla\cdot\frac{1}{2} \Re\{\hat{\mathbf{E}} \times \hat{\mathbf{H}}^{*}\}=\nabla\cdot\Re\{\hat{\mathbf{S}}\}=-\bar{q}_{e}(\omega),
\end{equation}
where $\langle\mathbf{S}\rangle$ is the time-averaged Poyinting vector and $\bar{q}_{e}$ is the time-averaged electric power dissipation density
\begin{equation}\label{Eq:FreqPowerDissipation}
\bar{q}_{e}(\omega)= \frac{1}{2}\varepsilon_{0}\,\omega \,\Im\left\{ \varepsilon(\omega) \right\}|\hat{\mathbf{E}}|^{2}.    
\end{equation}

\section{Density-based TopOpt using the FDTD method}\label{Sec:AdjointMethod}
In one of our previous works~\cite{GedeonDissip}, we presented a time-domain adjoint method based on the CCPR-ADEs for density-based topology optimization and explained its implementation in a FDTD framework. The term ``density-based'' refers to a particular structural optimization approach where a continuous density function~$\rho(\mathbf{r})$ in space is optimized (instead of varying the physical material layout directly)~\cite{Bendsoe2004}. The introduced density variable~\mbox{$\rho\in[0,1]$} allows for a smooth and differentiable interpolation between material-characteristic parameters of design and background media~(such as the permittivity $\varepsilon$). Its distribution in space can be iteratively updated based on the gradient information of a given objective function with respect to the density. The adjoint method allows for the computation of the full gradient information by performing only two simulations per iteration step - a \textit{forward} and an \textit{adjoint} simulation, which only differ in the source of excitation of the (photonic) system. As the optimization progresses, the density distribution gradually converges to a binary structure, where values of $\rho=1$ correspond to the design material and $\rho=0$ to the background material, defining the final optimized structure. To prevent a convergence to checkerboard patterns or the density from remaining ``grayish" (not binary), the density is filtered and projected at each iteration step~\cite{Wang2011OnPM}. It additionally allows one to enforce geometric constraints such as a minimum feature size of the device, which is significant for manufacturing~\cite{Piggott2017, Hammond:21}.

The material interpolation for the particular Silicon metasurface optimization in the main article relies on a \textit{linear} interpolation of the permittivities $\varepsilon_{\text{Si}}(\omega)$ and air $\varepsilon_{\text{Air}}=1$, 
\begin{equation}\label{Eq:InterpolationSpecific}
\varepsilon(\rho,\omega):= (1-\rho)\,\varepsilon_{\text{Air}} + \rho\, \varepsilon_{\text{Si}}(\omega).    
\end{equation}
To keep generality and for the sake of a simpler notation hereafter, we denote the background material with index $i=1$ and design material with index $i=2$ and assume that both media are described by the CCPR model~(\ref{Eq:CCPR_isotropic}). The Eq.~(\ref{Eq:InterpolationSpecific}) can then be rewritten as
\begin{equation}\label{Eq:InterpolationIndexed}
\varepsilon(\rho,\omega):= \sum_{i=1}^{2}\kappa^{(i)}(\rho)\,\varepsilon^{(i)},  
\end{equation}
where $\kappa^{(1)}(\rho):=1-\rho$ and $\kappa^{(2)}(\rho):= \rho$.

This density interpolation is then also reflected in the time-domain Maxwell's equations~(\ref{Eq:AuxFields_timedep}) and in the expression for the instantaneous electric power dissipation density~(\ref{Eq:InstantanousDissipation})~\cite[Sec.~III.A]{GedeonDissip}. Considering an incident pulse injected outside the design domain $\Omega$, the (density-interpolated) version of the Maxwell's equation~(\ref{Eq:AuxFields_timedep}) represent the \textit{forward system}, where the time-averaged power dissipation over a volume $\Omega_{\mathrm{o}}\subseteq\Omega$ is measured; this is the objective function for maximizing broadband absorption from the main article, where $\Omega_{\mathrm{o}}\equiv\Omega$. The corresponding \textit{adjoint system} is identical to the forward system except for the source of excitation. In this case, the (multiple) adjoint sources depend on the time-reversed forward fields and are injected into the volume $\Omega_{\mathrm{o}}$, 
\begin{align}
\mathbf{S}_{E}=&\; 2\,T^{-1}\sigma(\rho)\overleftarrow{\mathbf{E}}, \\[4pt]
\mathbf{S}_{\partial_{\tau}Q_{p}}^{(i)}=&\;2 \,T^{-1} \partial_{\tau}\overleftarrow{\mathbf{Q}}_{p}^{(i)},\;i=1,2;\; \forall p\, \in \{1, \ldots, P^{(i)}\}
\end{align}
where $T$ is the observing time period and ``$\overleftarrow{}$'' denotes the time-reversal.

\subsection{FDTD discretization}\label{Sec:FDTDdiscret}
\begin{figure}[h]
	\centering
	\includegraphics[width=0.8\linewidth]{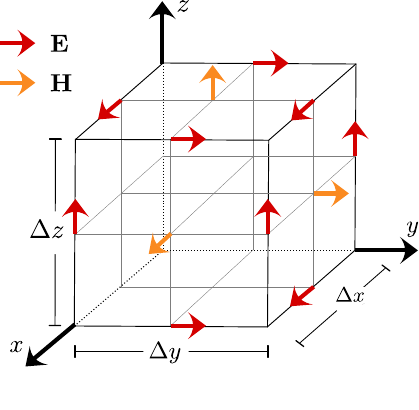}
	\caption{A single Yee cell in 3D with volume~$\Delta x\Delta y \Delta z$. The fields' components are distributed on a staggered grid, i.e., the electric (and auxiliary) field components are located centered and parallel to the cell edges, while the magnetic field components are located centered and parallel to the cell faces.}
	\label{fig:Yee}
\end{figure}

The gradient expression of the objective with respect to the density is an integral over the time period $T$ of adjoint and (time-reversed) forward fields~\cite[Sec.~III.A]{GedeonDissip}. The forward and adjoint equations can be discretized in space and time to provide an update scheme for the FDTD method. The time discretization of the forward system is identical to that presented in Sec.~\ref{Sec:CCPR_updateEqs}. The discretization of the adjoint equations, however, comes with special peculiarities where both adjoint source and fields in the gradient expression must be averaged with respect to the predetermined discretization of the forward equations~\cite[App.~C]{GedeonDissip}. Mapping these equations to the FDTD \textit{Yee} grid~\cite{Taflove}~(see Fig.~\ref{fig:Yee}) further requires a spatial discretization of the material distribution. We assigned the discretized density values to the locations of the electric field components~$E_k$, $k\in\{1,2,3\}$, within the staggered 3D~grid. If ``$n$" denotes the ``$n$-th Yee cell", the spatial position can be indexed by a tuple~$(k,n)$. By this definition and applying the time-discretization presented in our previous work~\cite[App.~C]{GedeonDissip}, the gradients can be represented and computed as~\cite[Sec.~III.C]{GedeonDissip}
\begin{equation}\label{Eq:GradientSuperposition}
\nabla_{\rho}F:= \mathcal{A}_{|\rho\in\Omega}+ \mathcal{B}_{|\rho\in\Omega_{\mathrm{o}}},
\end{equation}
where
\begin{subequations}\label{Eq:GradientsTimeReversal_Diss_FDTD}
\begin{align}
&\mathcal{A}_{\rho\in\Omega}:=\\[4pt]
-&\sum^{M}_{m=0}\varepsilon_{0}\mathrm{d}_{\rho}\varepsilon_{\infty}\mathbf{E}^{M-m}\cdot\left(\tilde{\mathbf{E}}^{m + \frac{1}{2}} - \tilde{\mathbf{E}}^{m - \frac{1}{2}}\right) \notag \\
 -&\sum^{M}_{m=0}\mathrm{d}_{\rho}\sigma\mathbf{E}^{M-m}\cdot\frac{1}{2}\left(\tilde{\mathbf{E}}^{m + \frac{1}{2}} + \tilde{\mathbf{E}}^{m - \frac{1}{2}}\right)\,\Delta t \notag\\
 -&\sum^{M}_{m=0}\sum_{i=1}^{2}\sum_{p=1}^{P^{(i)}}2\mathrm{d}_{\rho}\kappa^{(i)}\Re\left\{\mathbf{Q}^{M-m, (i)}_{p}\right\}\cdot\left(\tilde{\mathbf{E}}^{m + \frac{1}{2}} - \tilde{\mathbf{E}}^{m - \frac{1}{2}}\right), \notag \\[12pt]
 &\mathcal{B}_{\rho\in\Omega_{\mathrm{o}}}:=\\[4pt]
 &\sum^{M}_{m=0}\frac{\mathrm{d}_{\rho}\sigma}{T}\left(\mathbf{E}^{M-m}\right)^2\Delta t \notag \\
 +&\sum^{M}_{m=0}\sum_{i=1}^{2}\sum_{p=1}^{P^{(i)}}\mathrm{d}_{\rho}\kappa^{(i)}\Re\left\{\frac{\left(\mathbf{Q}^{M-1-m, (i)}_{p} - \mathbf{Q}^{M-m+1, (i)}_{p}\right)^2}{T\,2\,\Delta t\,\varepsilon_{0}\,c^{(i)}_{p}}\right\}. \notag
\end{align}
\end{subequations}
This formula is also applicable to anisotropic media. When this equation is used, however, the density values $\rho_{k,n}$ must be \textit{filtered} (taking their neighbourhood into account)~\cite{Bourdin01Filters}\,--\,otherwise the density might converge to three different component-wise structures that do not result in a physically meaningful topology. An extraction method for the final design based on the density distribution within the staggered 3D~Yee grid is presented in Ref.~\cite{Gedeon2023}. 

\section{Dissipation and power flow in the Ridges metasurface}\label{Sec:PowerFlow}
\begin{figure*}[!ht]
\centerline{\includegraphics[width=1\linewidth]{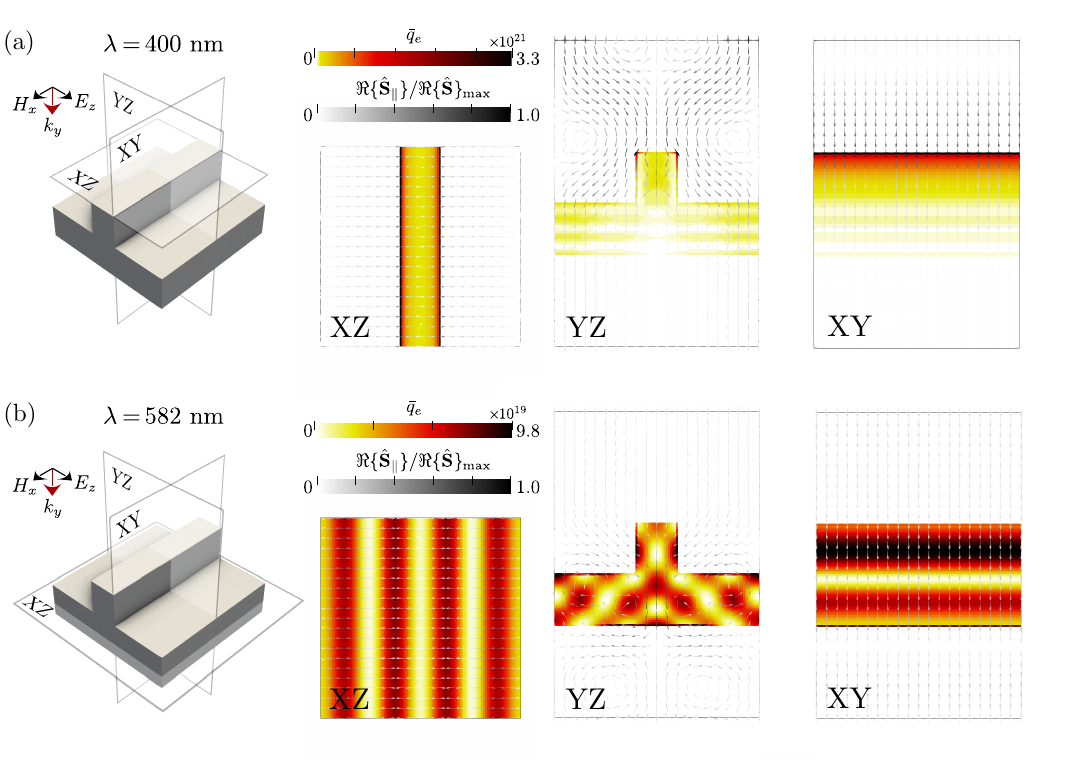}}
\caption{Cross-sections of the time-averaged electric power dissipation density $\bar{q}_e$~(\ref{Eq:FreqPowerDissipation}) and power flow for \mbox{$\lambda_{1}=400$\,nm}~(top) and \mbox{$\lambda_{2}=582$\,nm}~(bottom) within a unit cell of the Ridges metasurfaces presented in the main article~(Fig.~4). Both quantities are linked by the Eq.~\ref{Eq:PoyntingVector}. The dissipation is normalized to the power of the incident source (units:~$m^{-3}$). The power flow is represented by the Poynting vector's real part~$\Re\{\hat{\mathbf{S}}\}$, and the corresponding in-plane components for each cross-section are shown (normalized to the maximum). The full (3D) data of the time-harmonic electric field, time-averaged power dissipation, and Poynting vector are available in the dataset~\cite{dataset}.}
\label{fig:Ridges_poynting}
\end{figure*}

Fig.~\ref{fig:Ridges_poynting} illustrates cross-sections of both the time-averaged electric power dissipation density and  power flow of the Ridges structure presented in Fig.~4 of the main article, which are both linked by Eq.~(\ref{Eq:PoyntingVector}). The dissipation density is the negative of the divergence of the time-harmonic Poynting vector's real part (hereafter simply referred to as ``Poynting vector''); their spatial profiles provide information on where energy from the ($z$-polarized) incident plane wave is locally dissipated within the nanostructure. Due to the symmetry of the physical system, both quantities must be invariant along $x$ and have mirror symmetry along the (centered) $z$-axis. 
Figs.~\mbox{\ref{fig:Ridges_poynting}\,(a)\,\&\,(b)} show the profiles for the wavelengths~\mbox{$400$\,nm} and \mbox{$582$\,nm}, which we associated with a Mie-type and a QGM~resonance, respectively. The corresponding cross-sections were selected based on where the dissipation density particularly shows strong local hotspots contributing to the enhanced absorption.

For \mbox{$\lambda=400$\,nm}, we observe that the enhanced absorption comes primarily from the strong dissipation along the outer edges at the top of the Ridges~(see XZ and XY cross-sections). Within the bottom (Si) layer, we observe an interference pattern similar to Fabry-Perot leaky modes in a three-layer structure~\cite[SI, Sec.~1 (Fig.~S1)]{Huang2023} with only small perturbations of the Poynting vector directing in parallel to the $y$-axis~(see YZ cross-section). As this kind of interference is independent from the lattice periodicity, together with localized field confinement at the subwavelength (top) Ridges, motivates the attribution of a ``Mie-type''~resonance.
 A different picture yields the profiles for \mbox{$\lambda=582$\,nm}. We observe a delocalized distribution of multiple local hotspots within the entire structure, and a noticeable axial component of the Poynting vector along $z$. This mode profile, together with the associated absorption spike and spectral position of the first-order Rayleigh anomaly at \mbox{$\lambda^{\text{RA}} = 584$\,nm} (see main article), indicates a resonance of a guided mode.

\bibliographystyle{IEEEtran.bst}
\bibliography{main.bib}